\documentclass[preprint,showpacs,floatfix]{revtex4-1}
\usepackage{graphicx,psfrag,amsmath,amssymb,amsfonts,latexsym,color,epsf,graphpap}

\begin{document}

\title{Comment on: ``How the huge energy of quantum vacuum gravitates to drive the slow accelerating expansion of the Universe"}
\author{Francisco D. Mazzitelli and Leonardo G. Trombetta\footnote{Present address: Scuola Normale Superiore, Piazza dei Cavalieri 7, 56126, Pisa, Italy\\
INFN - Sezione di Pisa, 56200, Pisa, Italy}}
\affiliation{Centro At\'omico Bariloche and Instituto Balseiro, 
Comisi\'on Nacional de Energ\'\i a At\'omica, 8400 Bariloche, Argentina.}

\date{\today}

\begin{abstract} 
In a recent paper (Phys. Rev. D95, 103504 (2017)) it is argued  that, due to the fluctuations around its mean value, vacuum energy gravitates
differently from what previously assumed. As a consequence, the universe would accelerate with a small Hubble expansion rate, solving
the cosmological constant and dark energy problems. We point out here that the results  depend on the type of cutoff
used to evaluate the vacuum energy. In particular, they are not valid when one uses a covariant cutoff such that the zero point energy
density  is
positive definite.
\end{abstract}
\maketitle

In the traditional formulation of the cosmological constant problem, it is argued that the zero point energy density $\langle\rho\rangle$ associated to a quantum field is proportional to $\Lambda^4$, where $\Lambda$ is an ultraviolet cutoff, of the order of the Planck energy $E_{Planck}$. 
Assuming that the mean value of
the stress tensor of the quantum field is covariantly regularized, one has $\langle T_{\mu\nu}\rangle =-\langle\rho\rangle \, g_{\mu\nu}$, which corresponds to a cosmological constant of order $E_{Planck}^4$, about $120$ orders of magnitude larger than the observed one.

In Ref.\cite{Wang} it was pointed out that the energy-momentum tensor associated to a quantum massless field $\phi$  has very large fluctuations around its mean value. Therefore,  it is not correct to use $\langle T_{\mu\nu}\rangle $ as a source of the Einstein equations. When properly taken into account, these fluctuations lead to modified Einstein equations with a stochastic component. More concretely, for a metric of the form
\begin{equation}
ds^2=-dt^2+a^2(t,\mathbf x)\left(dx^2+dy^2+dz^2\right)
\end{equation}
the evolution equation for the scale factor $a(t,x)$ is that of a harmonic oscillator 
\begin{equation}
\ddot a +\Omega^2(t,\mathbf x) a=0\, \, ,\, \quad  \Omega^2(t,\mathbf x)=\frac{4\pi G}{3}\left(\rho+\sum_{i=1}^3P_i\right)=\frac{8\pi G}{3} \dot\phi^2 \, ,
\end{equation}
where $\rho=T_{00}$, $P_i=T_{ii}/a^2$. The quantity $\Omega^2$  is assumed to have a positive mean value $\langle \Omega^2\rangle $, of order $\Lambda^4$, and to have quasiperiodic stochastic fluctuations in a time scale
of order  $1/\Lambda$. Thus, due to parametric resonance, the scale factor has an exponential growth with a Hubble rate $H$ which is exponentially small in the limit $\Lambda\to \infty$, solving the cosmological constant problem. 

In this comment we would like to stress the following point: if the theory is regulated by a Lorentz invariant cutoff
in flat spacetime, then one has $\langle p\rangle =-\langle \rho\rangle $, and 
therefore $\langle \Omega^2\rangle =- 8\pi G\langle \rho\rangle/3 $. Moreover, if the cutoff is such that $\langle \rho\rangle > 0$, as usually assumed, then $\langle \Omega^2\rangle < 0$ and the whole picture of parametric resonance breaks down.

Let us be more explicit. Wang et al first computed $\langle \Omega^2\rangle$ in Minkowski spacetime using a non-invariant  cutoff $\Lambda$ such that $\vert \vec p\vert < \Lambda$, where $\vec p$ denotes the $3$-momentum of the modes of the scalar field. In this case, both $\langle\rho\rangle$ and $\langle p\rangle$ are positive definite and proportional to $\Lambda^4$. Note, however,  that for this particular cutoff one has $\langle p\rangle =\langle\rho\rangle/3$, breaking the Lorentz invariance of $\langle T_{\mu\nu}\rangle$. This has been noticed long ago in Ref.\cite{Ahkmedov}: a non-covariant cutoff cannot be used to estimate the vacuum contribution to the cosmological constant (see also Refs. \cite{Martin,Elias}). If, in spite of this, one accepts  the use of this cutoff, and assumes that the regularized quantities have physical meaning, then the conclusions of Ref.\cite{Wang} looks correct, although the initial problem is different: $\langle T_{\mu\nu}\rangle$ does not describe a cosmological constant  but a radiation fluid.

Wang et al also computed $\langle \Omega^2\rangle$ using a Lorentz invariant procedure inspired in Pauli-Villars method. The particular implementation 
of this method used in Ref.\cite{Wang} may give $\langle \Omega^2\rangle > 0$ (this is not completely clear from Eq.(195) in Ref.\cite{Wang}). Once more, if this were the case,  the analysis of the dynamical equation for the scale factor in Ref.\cite{Wang} would be correct, but at the price of regularizing the theory in such a way that $\langle \rho\rangle < 0$. Clearly, the use of this particular Lorentz invariant  cutoff would not be equivalent to the use of a cutoff in $3-$momentum space, since it produces a vacuum energy density with a different sign.

But the situation is even worse: the Pauli-Villars method produces ambiguous results for the polynomial divergences \cite{PV1,PV2,PV3}. Only the logarithmic divergences are univocally determined by the method. We illustrate this fact with an example discussed in Ref.\cite{PV1}. The regularized energy momentum tensor in Minkowski spacetime, using Pauli-Villars method,  is given by
\begin{equation}
\langle T_{\mu\nu}\rangle =-\frac{g_{\mu\nu}}{4}\sum_{i=0}^NC_iM_i^4\log\frac{M_i^2}{\mu^2}\, ,
\end{equation}
where the masses $M_i, i=1,2,...N$ are the regulators, $M_0=m$ is the mass of the field, $\mu$ is an arbitrary mass scale, $C_0=1$,  and the constants $C_i, i=1,2,...N$ satisfy
\begin{equation} 
\sum_{i=0}^N C_i\left(M_i^2\right)^p=0\, ,
\end{equation}
for $p=0,1,2$. Due to these conditions, the result is independent of the scale $\mu$.  In principle one can add an arbitrary number $N$ of regulator fields, the minimum being $N=3$ to satisfy the above constraints. It has been shown that, for $N=3$, the regularized version of the stress tensor produces a negative energy density. In the particular case $M_i=\Lambda$ one has
\begin{equation}
\langle T_{\mu\nu}\rangle =-\frac{g_{\mu\nu}}{128\pi^2}\left[-\Lambda^4+4 m^2\Lambda^2-m^4\left(3+2\log\frac{\Lambda^2}{m^2}\right)\right]\, .
\end{equation}
This particular approach gives $\langle \rho\rangle <0$ and  $\langle \Omega^2\rangle > 0$. However, when including additional regulator fields, there is a freedom in the choice of the constants $C_i$  that can be used to fix the value of the quartic divergence at an arbitrary value, even zero. The introduction of additional regulator fields, needed in curved spacetimes, also give arbitrary values for the polynomial divergences \cite{PV3}.   Other Lorentz invariant approaches, like inserting powers of  $\Lambda^2/(\Lambda^2-k^2-i\epsilon)$ in the divergent integrals, give only a quadratic divergence proportional to $m^2\Lambda^2$, and no quartic divergence  \cite{PV1}.

In summary,  if one regularizes the theory with the Pauli-Villars method,  $\langle\Omega^2\rangle$ is not positive definite,
and becomes negative when one imposes the ``physical" criterium that the vacuum energy density 
should be positive definite. In this case,   it is not true that the fluctuations of the stress tensor around its mean value lead to a solution 
of the cosmological constant problem, based on the parametric resonance mechanism proposed in Ref.\cite{Wang}. 
But most importantly, in light of the fact that different cutoffs produce ambiguous results  for the sign of $\langle\Omega^2\rangle$ and of the vacuum energy,
the physical meaning of the regularized quantities in this context
is doubtful.

One could wonder whether the fluctuations around a negative
$\langle\Omega^2\rangle$ could stabilize the upside-down harmonic oscillator in Eq. (2), through parametric stabilization 
\cite{stab}, softening the effect of the
cosmological constant. 
This seems difficult in the present model, given that the (quasi) frequency of the fluctuations is much smaller than  $(-\langle\Omega^2\rangle)^{1/2}$.
Moreover,
this mechanism would suffer the same ambiguities pointed out in this comment, that is,  it would depend on the 
particular implementation of the regularization method. It would be interesting
to analyze the eventual suppression of the cosmological constant by
parametric stabilization in the context of semiclassical stochastic gravity \cite{HuVer},
by studying 
the effect of noise in the renormalized Einstein-Langevin equation.

\section*{Acknowledgments}
This research was supported by ANPCyT, CONICET, and UNCuyo. We would like to thank D. L\'opez Nacir for discussions.

\end{document}